\documentclass[english]{article}
\usepackage[T1]{fontenc}
\usepackage[latin9]{inputenc}
\usepackage{float}
\usepackage{amsmath}
\usepackage{amssymb}
\usepackage{graphicx}

\makeatletter
\newcommand{\lyxaddress}[1]{
	\par {\raggedright #1
	\vspace{1.4em}
	\noindent\par}
}

\@ifundefined{showcaptionsetup}{}{%
 \PassOptionsToPackage{caption=false}{subfig}}
\usepackage{subfig}
\makeatother

\usepackage{babel}
\begin{document}
\title{Edge states of a three dimensional kicked rotor}
\author{Alexandra Bakman\thanks{sasha.bakman@gmail.com - Corresponding Author},
Hagar Veksler and Shmuel Fishman}
\maketitle

\lyxaddress{Technion - Israel Institute of Technology, Technion City, Haifa 3200004,
Israel}
\begin{abstract}
Edge localization is a fascinating quantum phenomenon. In this paper,
the underlying mechanism generating it is presented analytically and
verified numerically for a weakly kicked three-dimensional rotor.
Analogy to tight binding model in solid state physics is used. The
edge states result of the edge at zero angular momentum of the three-dimensional
kicked rotor.
\end{abstract}

\section{\label{sec:Introduction}Introduction}

Edge state localization is a fascinating quantum phenomenon and has
been explored extensively in various fields. Edge states have been
studied analytically and experimentally in photonic crystals \cite{Photonic_Crystals1,Numerical_Photonic_Crystal_Edge_States},
semiconductors \cite{Surface_States_SC_Experimental,Surface_States_Semiconductors2}
and topological insulators \cite{Surface_States_TI_Numerical,TI_Experimental,TI_Experimental2}.
In this paper, we study the edge states in angular momentum space
for a three-dimensional kicked rotor \cite{Blumel_Smilansky,Quantum_Resonances_Averbukh,Anderson_Localization_Molecules},
a paradigm system for studying quantum effects in a classically chaotic
system \cite{Book_kicked_rotor}. This system was numerically and
experimentally realized with planar molecules kicked by periodic short
microwave \cite{Theoretical_analysis_diatomic_molecules} and laser
\cite{Blumel_Smilansky,Quantum_Resonances_Averbukh,Rotor_permanent_dipole_moment_analysis}
pulses. The formation of the edge states was studied numerically \cite{Averbukh_paper}
for such a system excited at a fractional quantum resonance. In this
regime, the period of kicks is a rational fraction of the natural
period of the rotor \cite{Izrailev_and_Shepelyanski}.

The regime of quantum resonance \cite{Izrailev_and_Shepelyanski,Discovery_of_quantum_resonance}
has no analog in classical physics. In this regime, the energy of
the rotor grows quadratically with the number of kicks, since an initial
wavefunction explores higher angular momentum states as time progresses.
\cite{Anderson_Localization_Molecules}. However, since the angular
momentum quantum number cannot be negative in our system, it creates
quasienergy edge states, similar to the states on the surface of a
crystal. These states, centered near zero angular momentum, do not
acquire energy as time progresses.

In this paper, the mechanicm for creation of such states, resulting
of broken translational invariance symmetry, is presented. It uses
the fact that the $z$ component of angular momentum (in the standard
coordinate system) is zero. The translational invariance in angular
momentum is broken near in the vicinity of zero angular momentum.
In particular, we show that the wave number becomes imaginary, resulting
in exponential decay of such states. In the derivation we use perturbation
theory in order to understand in great detail the mechanism for creation
of the edge states. We believe that this mechanism is relevant also
in the non-perturbative regime, where some of the experiments are
performed \cite{Blumel_Smilansky,Quantum_Resonances_Averbukh,Anderson_Localization_Molecules,Theoretical_analysis_diatomic_molecules,Rotor_permanent_dipole_moment_analysis,Averbukh_paper}.

The system studied here is described by the Hamiltonian

\begin{equation}
H^{\prime}=\frac{\hbar^{2}}{2I}l\left(l+1\right)+\overline{P}\cos\left(\theta\right)\sum_{m=1}^{\infty}\delta\left(t^{\prime}-mT^{\prime}\right),\label{eq:Hamiltonian with dimensions}
\end{equation}

where $I$ is the moment of intertia, $\overline{P}$ the kick strength,
$T^{\prime}$ the period of the kicks and $\theta$ the planar angle
of the rotor. In dimensionless units $t=\frac{t^{\prime}}{T^{\prime}}$
, $H=\frac{H^{\prime}T^{\prime}}{\hbar},$ $\tau=\frac{\hbar T^{\prime}}{I}$
and $P=\frac{\overline{P}}{\hbar},$ this Hamiltonian becomes

\begin{equation}
H=\frac{1}{2}\tau l\left(l+1\right)+P\cos\left(\theta\right)\sum_{m=1}^{\infty}\delta\left(t-m\right).\label{eq:dimensionless Hamiltonian}
\end{equation}

Note that $\tau$ is the ratio between the driving period and the
natural period of the rotor \cite{Quantum_Localization_Shmuel}. Here,
we are looking at a rotor model where $m=0$.

In this paper, we aim to show analytically the creation of an edge
state in a regime of a fractional quantum resonance, therefore, $\tau$
has to be a rational fraction of $4\pi.$ For this purpose, we have
chosen
\begin{equation}
\tau=\frac{4\pi}{3}.\label{eq:tau}
\end{equation}

In order to describe the propagation of an initial wavefunction in
time, we write a transfer matrix $T$ which propagates the wavefunction
one kick forward in time

\begin{equation}
T\psi\left(t\right)=\psi\left(t+T\right).\label{eq:Definition of transfer matrix}
\end{equation}

We write this matrix as a product of two matrices

\begin{equation}
T=T_{kin}\cdot T_{P},\label{eq:Transfer matrix}
\end{equation}

where the kinetic and kick parts are

\begin{equation}
T_{kin}=e^{-i\frac{\tau}{2}l\left(l+1\right)}\label{eq:kinetic part}
\end{equation}

and

\begin{equation}
T_{P}=e^{-iP\cos\left(\theta\right)},
\end{equation}

respectively.

For this value of $\tau$, the kinetic part of the transfer matrix
(\ref{eq:kinetic part}) is periodic with period 3. The quasienergies
of the matrix $T_{kin}$ are

\begin{equation}
E_{1}=1\label{eq:E1_unperturbed}
\end{equation}

and

\begin{equation}
E_{2}=\alpha,\label{eq:E2_unperturbed}
\end{equation}

where 
\begin{equation}
\alpha=e^{i\frac{2\pi}{3}},\label{eq:alpha}
\end{equation}
 and $E_{1}$ is doubly degenerate. This degeneracy is lifted when
a correction to these energies due to the kicks is calculated.

For a free three-dimensional rotor, its eigenfunctions are spherical
harmonics, and $T_{P}$ is, for $m=0$ \cite{Blumel_Smilansky},

\begin{align}
T_{P,ll^{\prime}} & =\left\langle l,0\right|e^{-iP\cos\left(\theta\right)}\left|l^{\prime},0\right\rangle \label{eq:Transfer matrix calculation}\\
 & =\frac{\sqrt{\left(2l+1\right)\left(2l^{\prime}+1\right)}}{2}\int_{0}^{\pi}P_{l}\left(\cos\left(\theta\right)\right)P_{l^{\prime}}\left(\cos\left(\theta\right)\right)e^{-iP\cos\left(\theta\right)}\sin\left(\theta\right)d\theta,\nonumber 
\end{align}

where $P_{l}\left(\cos\left(\theta\right)\right)$ are Legendre polynomials.
Detailed discussion is found in Appendix \ref{sec:Transfer-matrix-calculation}.

We approximate the transfer matrix for weak kicking strength $P$
and large values of $l$ up to $P^{3}$ as

\begin{equation}
T=\begin{pmatrix}. & . & . & . & . & . & . & . & .\\
. & . & \alpha A & \alpha B & \alpha C & \alpha D & 0 & . & .\\
. & . & B & A & B & C & D & 0 & .\\
. & . & C & B & A & B & C & D & 0\\
. & . & \alpha D & \alpha C & \alpha B & \alpha A & \alpha B & \alpha C & \alpha D\\
. & . & 0 & D & C & B & A & B & C\\
. & . & . & 0 & D & C & B & A & B\\
. & . & . & . & . & . & . & . & .
\end{pmatrix},\label{eq:Transfer matrix large l with kicks}
\end{equation}

where
\begin{eqnarray}
A & \simeq & 1-\frac{P^{2}}{4}\label{eq:A}\\
B & \simeq & -i\frac{P}{2}+i\frac{P^{3}}{16}\label{eq:B}\\
C & \simeq & -\frac{P^{2}}{8}\label{eq:C}
\end{eqnarray}

and

\begin{equation}
D\simeq i\frac{P^{3}}{48}.
\end{equation}

The calculation can be found in Appendix \ref{sec:Transfer-matrix-calculation}.
This approximation holds for large values of $l,$ far from the edge.
We use the matrix (\ref{eq:Transfer matrix large l with kicks}) in
further calculations due to its periodicity. The correction for smaller
values of $l$ will manifest itself in the energy of the edge state.

The transfer matrix (\ref{eq:Transfer matrix large l with kicks})
is periodic, therefore in further calculations we use a solid state
tight binding system as an analog. This analog between a solid state
tight binding model \cite{Tight_Binding_first,Tight_Binding_Book,Bloch_Tight_Binding,Solid_State}
and a periodically kicked system is a convenient one, and was first
proposed in 1982 by Grempel, Fishman and Prange \cite{Shmuel_1982}.

The paper is organized as follows. In Sect. \ref{sec:Solid state tight binding analog}
the solid state analog of the transfer matrix is presented. In Sect.
\ref{sec:Quasienergies calculation} the quasienergies of a weakly
kicked rotor are approximated using perturbation theory. In Sect.
\ref{sec:Edge-state-formation} the quasienergy and wavenumber of
the edge state are calculated and compared to the numerical results.
The results are summarized and discussed in Sect. \ref{sec:Summary-and-discussion}.

\section{\label{sec:Solid state tight binding analog}Solid state tight binding
analog of the transfer matrix}

Since the transfer matrix (\ref{eq:Transfer matrix large l with kicks})
is periodic with period three, we use an analog of a solid state system
with three atoms in a unit cell. The eigenvectors of such a system
are

\begin{equation}
\phi=\sqrt{\frac{3}{N}}\sum_{l^{\prime}}\left(\alpha_{1}e^{ik\cdot\left(3l^{\prime}\right)}+\alpha_{2}e^{ik\cdot\left(3l^{\prime}+1\right)}+\alpha_{3}e^{ik\cdot\left(3l^{\prime}+2\right)}\right).\label{eq:eigenvectors solid state}
\end{equation}

where $N$ is the number of atoms in the crystal, $k$ is the wavenumber
and $\alpha_{1},\alpha_{2}$ and $\alpha_{3}$ are constants.

We therefore rewrite the transfer matrix as

\begin{equation}
T_{solid}=\begin{pmatrix}\alpha A+\alpha D\left(e^{3ik}+e^{-3ik}\right) & \alpha Be^{ik}+\alpha Ce^{-2ik} & \alpha Be^{-ik}+\alpha Ce^{2ik}\\
Be^{-ik}+Ce^{2ik} & A+D\left(e^{3ik}+e^{-3ik}\right) & Be^{ik}+Ce^{-2ik}\\
Be^{ik}+Ce^{-2ik} & Be^{-ik}+Ce^{2ik} & A+D\left(e^{3ik}+e^{-3ik}\right)
\end{pmatrix}.\label{eq:Transfer matrix solid state}
\end{equation}

A detailed calculation is found in Appendix \ref{sec:Quasienergy-calculation-for-a-solid-state-analog-system}.

\section{\label{sec:Quasienergies calculation}Calculating the quasienergies
using perturbation theory}

The eigenvalues of the matrix (\ref{eq:Transfer matrix solid state})
are of the form $E=\exp\left(i\omega\right),$ where $\omega$ are
the quasienergies of the rotor. After calculating $\det\left(T_{solid}-I\cdot E\right)$
and equating to zero, the following characteristic equation for the
energies is obtained

\begin{eqnarray}
E^{3}-E^{2}\left(1-\frac{P^{2}}{4}+i\frac{P^{3}}{48}\gamma\right)\left(\alpha+2\right)\label{eq:Energy equation}\\
-E\Bigg(\left(2\alpha+1\right)\left(-1+\frac{P^{2}}{4}\right)+\alpha i\frac{P^{3}}{24}\gamma+i\frac{P^{3}}{48}\gamma\Bigg)-\alpha & = & 0\nonumber 
\end{eqnarray}

Using (\ref{eq:Energy equation}), we calculate the correction to
(\ref{eq:E1_unperturbed}) and (\ref{eq:E2_unperturbed}) for small
$P$ using perturbation theory. The resulting energies are, up to
the third order in $P,$

\begin{equation}
E_{1}\simeq1\pm i\frac{1}{2}P+\frac{3\left(1+\alpha^{*}\right)\gamma-12\alpha^{*}}{48\left(\alpha^{*}-1\right)}P^{2}\label{eq:E1 with correction}
\end{equation}

and

\begin{equation}
E_{20}\simeq\alpha+\frac{\left(\alpha+1\right)}{4\left(\alpha^{*}-1\right)}P^{2}-\frac{iP^{3}\alpha\gamma}{48},\label{eq:E2 with correction}
\end{equation}

where

\begin{equation}
\gamma=e^{-3ik}+e^{3ik}.\label{eq:gamma}
\end{equation}

Detailed calculation can be found in Appendix \ref{sec:Quasienergy-calculation-for-a-solid-state-analog-system}.

Comparison between the numerical quasienergies for a three-dimensional
rotor and a theoretical result from (\ref{eq:E1 with correction})
is shown in Fig. \ref{fig:E vs. Etheory}. The edge state appears
at $k=0$ and its theoretical quasienergy is given by (\ref{eq:Edge state energy}).
Note that the edge state energy does not appear in the theoretical
graph, because the calculation was made for a periodic matrix, which
fits the transfer matrix for large values of $l,$ far from $l=0.$ 

Eq. (\ref{eq:E1 with correction}) yields the following dispersion
relation

\begin{equation}
\gamma=\frac{16\left(\alpha^{*}-1\right)}{\left(\alpha^{*}+1\right)P^{2}}\left(E+\frac{iP}{2}-1\right)+\frac{4\alpha^{*}}{1+\alpha^{*}}\label{eq:Dispersion relation}
\end{equation}

\section{\label{sec:Edge-state-formation}Edge state formation}

The edge state in a three-dimensional rotor system is formed at $l=0$,
where $l$ stands for angular momentum, because only its non-negative
values are allowed. This edge results in an eigenstate centered around
$l=0$ and whose decay is a function of the imaginary part of the
wavenumber $k.$

\subsection{Calculating $E_{edge}$}

We calculate the quasienergy of the edge state as an eigenvalue of
a $2\times2$ matrix $T_{0}$ that is a subset of the transfer matrix
$T$ close to $T_{00}$

\begin{equation}
T_{0}=\begin{pmatrix}T_{00} & T_{01}\\
T_{10} & T_{11}
\end{pmatrix},\label{eq:Small transfer matrix}
\end{equation}

this is because although the edge state is centered around $l=0,$
it has a ``tail'' which involves $l=1.$ In this matrix (\ref{eq:Small transfer matrix})
we also take into account the correction for small $l.$ Using (\ref{eq:Transfer matrix calculation})
yields

\begin{equation}
T_{0}=\begin{pmatrix}T_{00} & T_{01}\\
\alpha T_{01} & T_{11}
\end{pmatrix},
\end{equation}

where

\begin{align}
T_{00} & =1-\frac{P^{2}}{6},\\
T_{01} & =-\frac{iP}{\sqrt{3}}+\frac{iP^{3}}{10\sqrt{3}}
\end{align}

and

\begin{equation}
T_{11}=\alpha\left(1-\frac{3P^{2}}{10}\right),
\end{equation}

respectively. 

The energy of the edge state is one of the eigenvalues of (\ref{eq:Small transfer matrix})
and it is, up to $P^{3},$

\begin{equation}
E_{edge}=\frac{1}{2}\left(1+\alpha-P^{2}\cdot\frac{5+9\alpha}{30}+\sqrt{\left(1-\alpha\right)^{2}-\frac{P^{2}}{15}\left(9\alpha^{2}+6\alpha+5\right)}\right).\label{eq:Edge state energy}
\end{equation}

\begin{figure}
\subfloat[]{\includegraphics[scale=0.4]{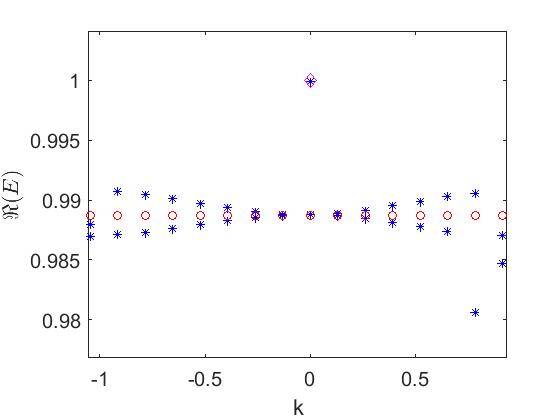}}\subfloat[]{\includegraphics[scale=0.4]{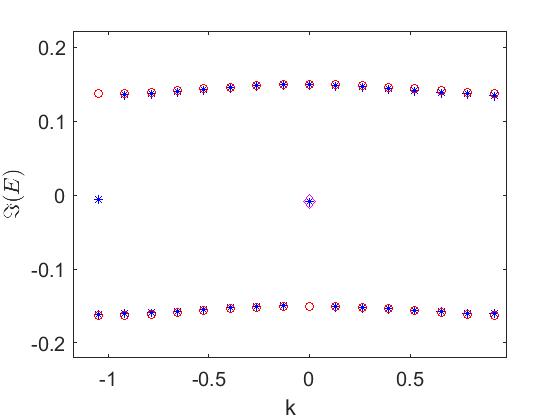}

}

\subfloat[]{\includegraphics[scale=0.4]{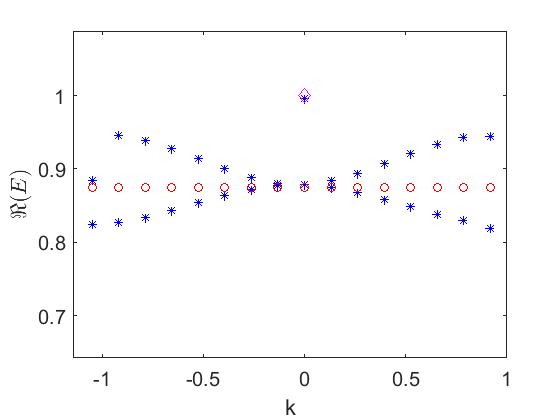}

}\subfloat[]{\includegraphics[scale=0.4]{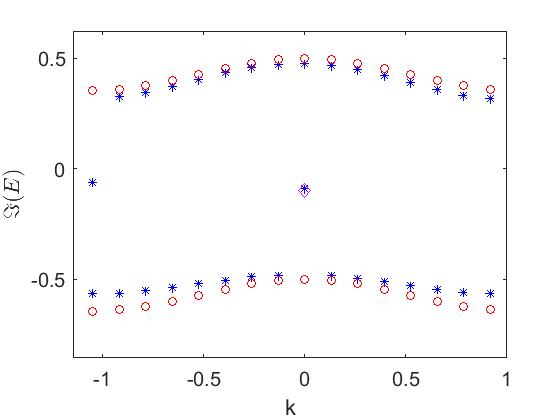}

}\caption{\label{fig:E vs. Etheory}Comparison between the real and imaginary
parts of the quasienergy numerically found using the transfer matrix
(\ref{eq:Transfer matrix}) for a three-dimensional kicked rotor and
a theoretical result (\ref{eq:E2 with correction}) for $\tau=4\pi/3$
and $P=0.3$ ((a),(b)) and $P=1$ ((c),(d)). The numerical result
is shown by asterisks, the theoretical result based on (\ref{eq:dimensionless Hamiltonian})
is shown by circles and the theoretical energy of the edge state (\ref{eq:Edge state energy})
is shown by a diamond.}
\end{figure}

A detailed calculation is found in Appendix \ref{sec:Calculating-the-energy-of-the-edge-state}.
This energy is shown in Fig. \ref{fig:E vs. Etheory} for $P=0.3$
and $P=1.$

\subsection{Calculating $k_{edge}$}

In order to calculate the wavenumber corresponding to (\ref{eq:Edge state energy}),
we substitute (\ref{eq:Edge state energy}) into (\ref{eq:Dispersion relation})
and use

\begin{equation}
k_{1,2}=\frac{i}{3}\ln\left(\frac{\gamma_{edge}}{2}\pm\sqrt{\frac{\gamma_{edge}^{2}}{4}-1}\right)\label{eq:k edge state}
\end{equation}

where

\begin{equation}
\gamma_{edge}=e^{3ik_{edge}}+e^{-3ik_{edge}}.
\end{equation}

The absolute value of the edge state $\psi_{edge}$ as function of
$l$ is shown in Fig. \ref{fig:Edge state decay 3D} below. Notice
that it is large at zero angular momentum.

\begin{figure}[H]
\centering{}\subfloat[]{\includegraphics[scale=0.4]{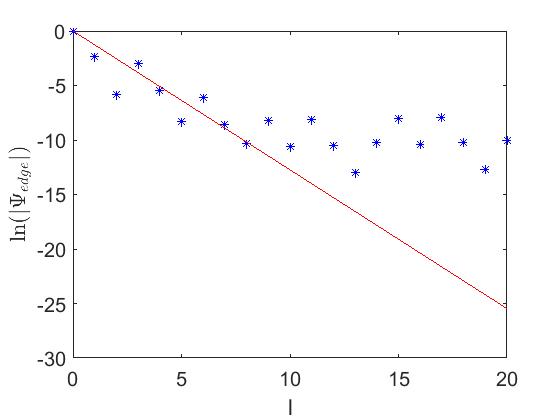}}\subfloat[]{\includegraphics[scale=0.4]{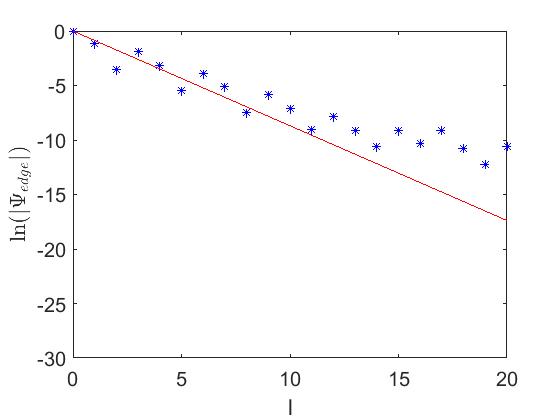}}\caption{\label{fig:Edge state logarithm}Comparison between the logarithm
of the quasienergy state $\ln\left|\psi_{edge}\right|$ found numerically
using the Hamiltonian (\ref{eq:dimensionless Hamiltonian}) and $\ln\left|\psi_{edge}\right|$
of (\ref{eq:edge state decay}) for (a) $P=0.3$, (b) $P=1$ and $\tau=4\pi/3$
for a three dimensional kicked rotor. The numerical result is shown
by asterisks and the theoretical result by a continuous line.}
\end{figure}

Since $k_{edge}$ has an imaginary part, it causes the corresponding
state to decay exponentially as function of $l.$ Hence, the absolute
value of the edge state can be approximated as
\begin{equation}
\left|\psi_{edge}\right|\varpropto e^{-\text{\ensuremath{\Im\left(k_{edge}\right)}}l}.\label{eq:edge state decay}
\end{equation}
The comparison between the absolute value of the edge state and a
model state with this decay rate is shown in Fig. \ref{fig:Edge state decay 3D}
for $P=0.3$ and $P=1$. The logarithm of $\left|\psi_{edge}\right|$
is compared to the theoretical slope according to (\ref{eq:edge state decay})
in Fig. \ref{fig:Edge state logarithm}.

We repeated the calculation for $P=2$ and $P=3$ and found considerable
deviation from the numerical result for the energies. However, the
edge states presented reasonable agreement with the numerical results.

\begin{figure}
\centering{}\subfloat[]{\includegraphics[scale=0.4]{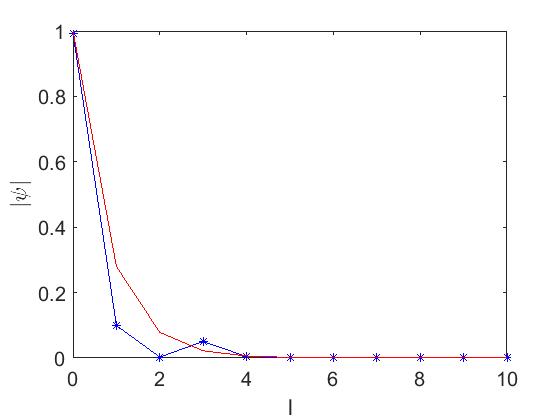}

}\subfloat[]{\includegraphics[scale=0.4]{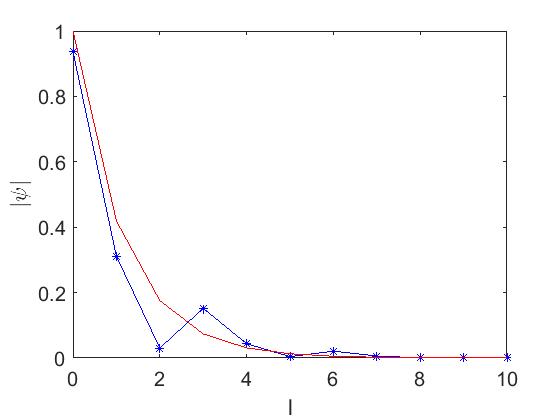}

}\caption{\label{fig:Edge state decay 3D}Comparison between the quasienergy
state$\left|\psi_{edge}\right|$ found numerically using the transfer
matrix (\ref{eq:Transfer matrix})and $\left|\psi_{edge}\right|$
of (\ref{eq:edge state decay}) for (a) $P=0.3$, (b) $P=1$ and $\tau=4\pi/3$
for a three dimensional kicked rotor. The numerical result is shown
by asterisks and the theoretical result by a continuous line.}
\end{figure}

\section{\label{sec:Summary-and-discussion}Summary and discussion}

In a three-dimensional rotor system studied here, edge states appear
due to the edge in angular momentum at $l=0$ (see Eq. (\ref{eq:Transfer matrix calculation})).
An initial wavevector located near $l$=0 will stay there as time
progresses. Near this region, the transfer matrix (\ref{eq:Transfer matrix large l with kicks})
becomes dependent on $l$ while for large values of $l$ it is similar
to the two dimensional rotor matrix (and independent of $l$).

We analytically calculated the energy and decay rate of the edge states
for the three dimensional rotor system using an analog tight binding
model, based on the periodicity of the transfer matrix. The energy
of the edge state was found along with its exponential decay rate,
and a comparison between the theoretical and numerical results was
shown in Figs. \ref{fig:Edge state decay 3D} and \ref{fig:Edge state logarithm}.
In this paper, it was demonstrated analytically in great detail how
edge states in angular momentum, that were found numerically in the
past \cite{Averbukh_paper}, are generated. We calculated the edge
states also for $P=2$ and $P=3$ and found reasonable agreement with
the numerical results. Therfore, we believe this mechanism is valid
also in the non-perturbative regime, where most experiments are done
\cite{Blumel_Smilansky,Quantum_Resonances_Averbukh,Anderson_Localization_Molecules,Theoretical_analysis_diatomic_molecules,Rotor_permanent_dipole_moment_analysis,Averbukh_paper}.
This result is expected to motivate future research, in particular
of the non-perturbative regime.

\section*{Acknowledgements}

We thank Prof. Ilya Averbukh for fruitful and highly informative discussions.
We would like to acknowledge partial support of Israel Science Foundation
(ISF) Grant No. 931/16.

\section*{Author contribution statement}

All authors contributed equally to the research and presentation in
the paper.

\appendix

\section{\label{sec:Transfer-matrix-calculation}Transfer matrix calculation
for three-dimensional rotor}

\numberwithin{equation}{section} 

In order to calculate the transfer matrix elements for the three dimensional
rotor, we use the fact that for $m=0$ the spherical harmonics can
be written using Legendre polynomials as

\begin{equation}
Y_{\theta}^{0}=\sqrt{\frac{2l+1}{4\pi}}P_{l}\left(\cos\left(\theta\right)\right),
\end{equation}

where $P_{l}$ are Legendre polynomials. We also use the following
formula for integration of Legendre polynomials \cite{Legendre_integral}

\begin{equation}
\int_{-1}^{1}P_{k}\left(x\right)P_{l}\left(x\right)P_{m}\left(x\right)dx=2\begin{pmatrix}k & l & m\\
0 & 0 & 0
\end{pmatrix}^{2}\label{eq:triple Legendre integral}
\end{equation}

where

\begin{equation}
\begin{pmatrix}k & l & m\\
0 & 0 & 0
\end{pmatrix}=\left(-1\right)^{s}\sqrt{\frac{\left(2s-2k\right)!\left(2s-2l\right)!\left(2s-2m\right)!}{\text{\ensuremath{\left(2s+1\right)}!}}}\cdot\frac{s!}{\left(s-k\right)!\left(s-l\right)!\left(s-m\right)!}
\end{equation}

for $2s$ even, and zero otherwise, where

\begin{equation}
2s=k+l+m,
\end{equation}

and calculate the integrals corresponding to each diagonal up to the
third diagonals to the right and left (since the calculation is up
to $P^{3}$) , using the relation

\begin{equation}
\cos^{3}\left(\theta\right)=\frac{3}{5}P_{1}\left(\cos\left(\theta\right)\right)+\frac{2}{5}P_{3}\left(\cos\left(\theta\right)\right).\label{eq:cos3}
\end{equation}

For example, for the main diagonal we have

\begin{align}
\left\langle l,0\right|e^{-iP\cos\left(\theta\right)}\left|l,0\right\rangle  & \simeq\left\langle l,0\right|1-iP\cos\left(\theta\right)-\frac{P^{2}}{2}\cos^{2}\left(\theta\right)+\frac{i}{6}P^{3}\cos^{3}\left(\theta\right)\left|l,0\right\rangle 
\end{align}

We know that

\begin{align}
-iP\left\langle l,0\right|\cos\left(\theta\right)\left|l,0\right\rangle  & =-iP\frac{2l+1}{2}\int_{0}^{\pi}P_{l}^{2}\left(\cos\left(\theta\right)\right)\cos\left(\theta\right)\sin\left(\theta\right)d\theta\\
 & =-iP\frac{2l+1}{2}\int_{-1}^{1}P_{l}^{2}\left(x\right)xdx=0\nonumber 
\end{align}

because in the integral we have an antisymmetric function integrated
over a symmetric interval.

The integral for $P^{2}$ yields

\begin{align}
-\frac{P^{2}}{4}\left(2l+1\right)\int_{0}^{\pi}P_{l}^{2}\left(\cos\left(\theta\right)\right)\cos^{2}\left(\theta\right)\sin\left(\theta\right)d\theta & =-\frac{P^{2}}{4}\left(2l+1\right)\int_{-1}^{1}P_{l}^{2}\left(x\right)x^{2}dx\\
 & =-\frac{P^{2}}{2}\frac{2l^{2}+2l-1}{\left(2l-1\right)\left(2l+3\right)}\nonumber 
\end{align}

and the integral for $P^{3}$ yields, using (\ref{eq:cos3}) and (\ref{eq:triple Legendre integral}),

\begin{align}
\frac{iP^{3}}{6}\int_{-1}^{1}P_{l}^{2}\left(x\right)x^{3}dx & =\frac{iP^{3}}{10}\int_{-1}^{1}P_{l}^{2}\left(x\right)P_{1}\left(x\right)dx+\frac{iP^{3}}{15}\int_{-1}^{1}P_{l}^{2}\left(x\right)P_{3}\left(x\right)dx\\
 & =\frac{iP^{3}}{5}\begin{pmatrix}l & l & 1\\
0 & 0 & 0
\end{pmatrix}^{2}+\frac{2iP^{3}}{15}\begin{pmatrix}l & l & 3\\
0 & 0 & 0
\end{pmatrix}^{2}=0\nonumber 
\end{align}

because $2s$ is odd in both expressions.

The results for each diagonal are as follows. For the main diagonal
the result is

\begin{equation}
T_{main}=1-\frac{P^{2}}{2}\cdot\frac{2l^{2}+2l-1}{\left(2l-1\right)\left(2l+3\right)}.\label{eq:T_main}
\end{equation}

For the first diagonal to the right and left it is

\begin{equation}
T_{1st,right}=-iP\cdot\frac{l+1}{\sqrt{\left(2l+1\right)\left(2l+3\right)}}+\frac{iP^{3}}{20}\left[\frac{2l+2}{\sqrt{\left(2l+1\right)\left(2l+3\right)}}+\frac{l\left(2l+2\right)\left(l+2\right)}{\left(2l-1\right)\left(2l+5\right)\sqrt{\left(2l+1\right)\left(2l+3\right)}}\right]\label{eq:T_1st_right}
\end{equation}

and

\begin{equation}
T_{1st,left}=-iP\cdot\frac{l}{\sqrt{\left(2l+1\right)\left(2l-1\right)}}+i\frac{P^{3}}{20\sqrt{\left(2l+1\right)\left(2l-1\right)}}.\left[2l+\frac{2l\left(l-1\right)\left(l+1\right)}{\left(2l-3\right)\left(2l+3\right)}\right].\label{eq:T_1st_left}
\end{equation}

For the second diagonal to the right and to the left it is

\begin{equation}
T_{2nd,right}=-\frac{P^{2}}{2}\cdot\frac{\left(l+1\right)\left(l+2\right)}{\left(2l+3\right)\sqrt{\left(2l+1\right)\left(2l+5\right)}}\label{eq:T_2nd_right}
\end{equation}

and

\begin{equation}
T_{2nd,left}=-\frac{P^{2}}{2}\cdot\frac{l\left(l-1\right)}{\left(2l-1\right)\sqrt{\left(2l+1\right)\left(2l-3\right)}}.\label{eq:T_2nd_left}
\end{equation}

For the third diagonal to the right and to the left it is

\begin{equation}
T_{3rd,right}=i\frac{P^{3}}{48}\cdot\frac{\left(2l+2\right)\left(2l+4\right)\left(2l+6\right)}{\left(2l+3\right)\left(2l+5\right)\sqrt{\left(2l+1\right)\left(2l+7\right)}}\label{eq:T_3rd_right}
\end{equation}

and

\begin{equation}
T_{3rd,left}=\frac{iP^{3}}{48}\cdot\frac{2l\left(2l-2\right)\left(2l-4\right)}{\left(2l-3\right)\left(2l-1\right)\sqrt{\left(2l+1\right)\left(2l-5\right)}}.\label{eq:T_3rd_left}
\end{equation}

All of these expressions reach constants for large values of $l.$
These constants are

\begin{equation}
T_{main}\simeq1-\frac{P^{2}}{4}=A
\end{equation}

\begin{equation}
T_{1st,right}=T_{1st,left}\simeq-i\frac{P}{2}+\frac{iP^{3}}{16}=B
\end{equation}

\begin{equation}
T_{2nd,right}=T_{2nd,left}\simeq-\frac{P^{2}}{8}=C
\end{equation}

and

\begin{equation}
T_{3rd,right}=T_{3rd,left}\simeq i\frac{P^{3}}{48}=D,
\end{equation}

respectively.

\section{\label{sec:Quasienergy-calculation-for-a-solid-state-analog-system}Quasienergy
calculation for a solid state analog system}

The transfer matrix for the solid state analog is calculated as follows.

We can write $\phi$ of (\ref{eq:eigenvectors solid state}) as three
orthogonal vectors, and the new transfer matrix is built from casting
(\ref{eq:Transfer matrix large l with kicks}) on these vectors. One
example of this is:

\begin{align}
T_{s,11} & =\frac{1}{N}\begin{pmatrix}1 & 0 & 0 & e^{-3ik} & 0 & 0 & e^{-6ik} & 0 & . & .\end{pmatrix}\begin{pmatrix}\alpha A & \alpha B & \alpha C & \alpha D & 0 & 0 & 0 & 0 & . & .\\
B & A & B & C & D & 0 & 0 & 0 & . & .\\
C & B & A & B & C & D & 0 & 0 & . & .\\
\alpha D & \alpha C & \alpha B & \alpha A & \alpha B & \alpha C & \alpha D & 0 & . & .\\
0 & D & C & B & A & B & C & D & . & .\\
0 & 0 & D & C & B & A & B & C & . & .\\
0 & 0 & 0 & \alpha D & \alpha C & \alpha B & \alpha A & \alpha B & . & .\\
0 & 0 & 0 & 0 & D & C & B & A & . & .\\
. & . & . & . & . & . & . & . & . & .
\end{pmatrix}\\
\cdot & \begin{pmatrix}1\\
0\\
0\\
e^{3ik}\\
0\\
0\\
e^{6ik}\\
0\\
.\\
.
\end{pmatrix}\nonumber \\
 & =\frac{1}{N}\begin{pmatrix}1 & 0 & 0 & e^{-3ik} & 0 & 0 & e^{-6ik} & 0 & . & .\end{pmatrix}\begin{pmatrix}\alpha A+\alpha De^{3ik}\\
B+Ce^{3ik}\\
C+Be^{3ik}\\
\alpha D+\alpha Ae^{3ik}+\alpha De^{6ik}\\
Be^{3ik}+Ce^{6ik}\\
Ce^{3ik}+Be^{6ik}\\
\alpha De^{3ik}+\alpha Ae^{6ik}\\
Be^{6ik}\\
.\\
.
\end{pmatrix}\nonumber \\
 & =\alpha A+\alpha D(e^{3ik}+e^{-3ik}).\nonumber 
\end{align}

\section{\label{sec:Calculating-the-energy-of-the-edge-state}Calculating
the energy of the edge state}

In order to calculate the energy of the edge state, we calculate the
matrix (\ref{eq:Small transfer matrix}). 

From (\ref{eq:T_main}) we calculate $T_{00}$ by plugging in $l=0$

\begin{align}
T_{00} & =T_{main}\left(l=0\right)=1-\frac{P^{2}}{6}
\end{align}

and $T_{11}$by using $l=1$ and multiplying by $\alpha$ due to the
kinetic part of the matrix

\begin{equation}
T_{11}=\alpha T_{main}\left(l=1\right)=\alpha\left(1-\frac{3P^{2}}{10}\right).
\end{equation}

From (\ref{eq:T_1st_right}) we calculate $T_{01}$ by plugging in
$l=0$

\begin{equation}
T_{01}=T_{1st,right}\left(l=0\right)=-\frac{iP}{\sqrt{3}}+\frac{iP^{3}}{10\sqrt{3}}
\end{equation}

From (\ref{eq:T_1st_left}) we calculate $T_{10}$ by plugging in
$l=1$ and multiplying by $\alpha$ due to the kinetic part of the
matrix

\begin{equation}
T_{10}=\alpha T_{1st,left}\left(l=1\right)=\alpha\left(-\frac{iP}{\sqrt{3}}+\frac{iP^{3}}{10\sqrt{3}}\right)=\alpha T_{01}.
\end{equation}

The energy of the edge state $E_{edge}$ is one of the eigenvalues
of the matrix $T_{0}$. Calculating $\det\left(T_{0}-E\right)$ and
equating to zeros yields, up to $P^{3},$

\begin{align}
\left(1-\frac{P^{2}}{6}-E\right)\left(\alpha-\frac{3\alpha P^{2}}{10}-E\right)-\alpha\left(-\frac{iP}{\sqrt{3}}+\frac{iP^{3}}{10\sqrt{3}}\right)^{2} & =0\label{eq:determinant of T_0}\\
E^{2}-E\left(1+\alpha-P^{2}\frac{\left(5+9\alpha\right)}{30}\right)+\alpha\left(1-\frac{2P^{2}}{15}\right) & =0\nonumber 
\end{align}

This is a quadratic equation, whose solutions are

\begin{align}
E_{1,2} & =\frac{1+\alpha-P^{2}\frac{5+9\alpha}{30}\pm\sqrt{\left(1+\alpha-P^{2}\frac{\left(5+9\alpha\right)}{30}\right)^{2}-4\alpha\left(1-\frac{2P^{2}}{15}\right)}}{2}\\
 & \frac{1+\alpha-P^{2}\cdot\frac{5+9\alpha}{30}\pm\sqrt{\left(1-\alpha\right)^{2}-\frac{P^{2}}{15}\left(5+6\alpha+9\alpha^{2}\right)}}{2}\nonumber 
\end{align}

The solution for the edge state is with the plus sign before the square
root because it's an energy that belongs to the upper band and is
close in its value to $T_{00}.$

\bibliographystyle{IEEEtran}
\bibliography{Edge_States_References}

\end{document}